\documentclass[usenatbib]{mn2e}

\usepackage{aas_macros}
\usepackage{hyperref}
\usepackage{graphicx}
\usepackage{amsmath}
\usepackage{amssymb}
\usepackage[utf8]{inputenc}

\usepackage{lmodern}

\newcommand{\Mpcoverh}{~\mathrm{Mpc}/h}
\newcommand{\kpcoverh}{~\mathrm{kpc}/h}
\newcommand{\kpc}{~\mathrm{kpc}}
\newcommand{\hoverMpc}{~h.\mathrm{Mpc}^{-1}}
\newcommand{\dd}{{\, \mathrm{d}}}
\newcommand{\Rvir}{R_{\mathrm{vir}}}
\newcommand{\vect}[1]{\mathbf{#1}}
\newcommand{\Msun}{~M_{\odot}}
\newcommand{\Msunoverh}{~M_{\odot}/h}
\newcommand{\RAMSES}{{\sc ramses}}

\newcommand{\AbstractText}{
We study the dissipative effects of baryon physics on cosmic statistics at small scales using a cosmological simulation of a $(50 \Mpcoverh)^3$ volume of the universe.
The MareNostrum simulation was performed using the AMR code \RAMSES, and includes most of the physical ingredients which are part of the current theory of galaxy formation, such as metal-dependent cooling and UV heating, subgrid modelling of the ISM, star formation and supernova feedback.
We re-ran the same initial conditions for a dark matter only universe, as a reference point for baryon-free cosmic statistics.
In this paper, we present the measured small-scale amplification of $\sigma^2$ and $S_3$ due to baryonic physics and their interpretation in the framework of the halo model.
As shown in recent studies, the effect of baryons on the matter power spectrum can be accounted for at scales $k \lesssim 10 \hoverMpc$ by modifying the halo concentration parameter.
We propose to extend this result by using a composite halo profile, which is a linear combination of a NFW profile for the dark matter component, and an exponential disk profile mimicking the baryonic component at the heart of the halo.
This halo profile form is physically motivated, and depends on two parameters, the mass fraction $f_d$ of baryons in the disk, and the ratio $\lambda_d$ of the disk's characteristic scale to the halo's virial radius.
We find this composite profile to reproduce both the small-scale variance and skewness boosts measured in the simulation up to $k \sim 10^2 \hoverMpc$ for physically meaningful values of the parameters $f_d$ and $\lambda_d$.
Although simulations like the one presented here usually suffer from various problems when compared to observations, our modified halo model could be used as a fitting model to improve the determination of cosmological parameters from weak lensing convergence spectra and skewness measurements.
}

\newcommand{\KeywordsText}{
Gravitational lensing --
cosmological parameters --
Galaxy: disk --
Hydrodynamics --
large-scale structure of Universe
}

\newcommand{\AckText}{
The authors thankfully acknowledge the computer resources, technical expertise and assistance provided by the Barcelona Supercomputing Centre -- Centro Nacional de Supercomputaci\'on (\url{http://www.bsc.es/}).
The dark matter only re-run was carried out at the CEA CCRT computing centre (\url{http://www-ccrt.cea.fr/}).
We are grateful to J.~Fry and D.~Weinberg for useful discussions on the halo model.
We would also like to thank A.~Refregier and A.~Amara for their helpful insights and comments on the weak lensing aspects of this study.
}

\title{The effect of baryons on the variance and the skewness of the mass distribution in the universe at small scales}

\title[Baryons and statistics of the mass distribution] %
{The effect of baryons on the variance and the skewness of the mass distribution in the universe at small scales}

\author[T. Guillet, R. Teyssier and S. Colombi] %
{T. Guillet$^{1}$\thanks{E-mail: thomas.guillet@cea.fr}, R. Teyssier$^{1,2}$ and S. Colombi$^{3}$. \\
$^{1}$Service d'astrophysique, CEA/Saclay, F-91191 Gif-sur-Yvette, France \\
$^{2}$Universität Z\"urich,	Institute für Theoretische Physik, Winterthurerstrasse 190,	CH-8057 Z\"urich \\
$^{3}$Institut d'Astrophysique de Paris, CNRS UMR 7095 \& UPMC, 98bis, bd Arago, F-75014 Paris}

\begin{document}

\maketitle

\begin{abstract}
\AbstractText
\end{abstract}


\label{firstpage}

\section{Introduction}

One of the great challenges in modern cosmology is to understand the nature of dark energy, which is believed to dominate the energy budget in the universe ($\sim 70\%$) at low redshift \citep{Riess1998, Perlmutter1999, Spergel2007, Astier2006}.
Since the value of $\Omega_\Lambda$ directly impacts the rate of structure formation at recent epochs, the mass distribution and its time evolution bear the signature of the dark energy content of the universe.
Cosmic shear measurements provide an independent method of probing the total mass distribution at large scales.
Combined with photometric redshifts, it is even possible to extract the 3D matter distribution and reconstruct the matter power spectrum at different epochs.
Comparing these measurements to theoretical predictions will set strong constraints on the cosmological parameters \citep[e.g.,][]{Hu1999, Huterer2002, Amara2006, Albrecht2007}, 
and among them both the equation of state $w$ of dark energy and its possible evolution $w'$ with redshift.

The cosmic shear convergence power spectrum $P_{\kappa}$ depends on the total matter power spectrum $P$, which contains the information about $w$ and $w'$ through the growth rate of structures.
Extracting the dark energy equation of state from weak lensing signals therefore requires a good theoretical model for $P$, with a typical accuracy of a few percent up to angular scales of about 10' \citep[see for example][for a review]{Refregier2003,Bartelmann2001}.
Substantial work has been done to measure and predict the dark matter power spectrum from collisionless $N$-body simulations \citep[see, e.g.,][]{Efstathiou1981,Jenkins1998}.
Semi-analytic models for the dark matter power spectrum have also been proposed, reaching the percent level accuracy \citep{Hamilton1991,Jain1995,Peacock1996,Smith2003}.
While the distribution of total matter is likely to closely follow dark matter at large scales, dissipative physics is expected to modify the power spectrum at small scales, and therefore possibly interfere with weak lensing measurements. 

The interest for the effects of baryons on the convergence power spectrum has led to the development of semi-analytic halo models to estimate effect of cold \citep{White2004} and hot \citep{Zhan2004} gas on the total matter power spectrum.
More recently, numerical simulations have been carried out to complement those semi-analytical results \citep{Jing2006,Rudd2008}.
While the exact effect of the baryons differ quantitatively between different models, the models and simulations agree qualitatively on a boost of the total matter power spectrum due to cold baryons at small scales.
At $k \sim 10\hoverMpc$, this amplification has been found to be of around $10\%$ at $z=0$.
Our theoretical understanding of galaxy formation is however far from being complete.
Current numerical simulations that include various complex baryons physical processes suffer from the so-called \emph{overcooling problem} \citep{Blanchard1992,Cole1991}, with too many baryons condensing into gaseous and stellar disks when compared to observational constraints.
Statistical effects measured in Galaxy formation simulations, including the one used in the present work, are therefore likely overestimated.
If we can account for the effect of baryons at the required accuracy in this extreme case, real datasets will be probably even easier to deal with.

We still need a flexible and accurate tool to account for the effect of baryons on the statistical properties of the matter density field in a parametrised model.
The halo model has been developed in the last decade to meet these goals.
The halo model is based on the idea that the matter distribution in the universe can be described as a collection of individual halos, in which baryonic structures such as galaxies form \citep{Neyman1952, White1978}.
\citet{Scherrer1991} proposed a formalism to compute correlation functions of the continuous density field from a model of discrete virialized halos.
Since then, there has been notable contributions and refinements to the halo model approach, such as \citet{Ma2000,Seljak2000} and subsequent developments (see \citealt{Cooray2002} for a review in the context of large scale structure).

As the halo model has proved to be a successful framework for describing statistical properties of the dark matter density field in the non-linear regime, there has been also interest in extending it to baryons in the context of the Sunyaev-Zeldovich effect \citep{Refregier2002} and of the galaxy distribution \citep{Seljak2000}.
In previous studies, \citet{White2004} and \citet{Zhan2004} have used the halo model with a baryonic component to describe the effect of cold and hot gas respectively from a semi-analytical standpoint.
More recently, \citet{Rudd2008} have shown that the halo model can be used in a self-consistent way to describe the amplification of the power spectrum caused by baryons as measured in cosmological simulations.
They proposed to modify the concentration parameter mass dependence of the dark matter halos to account for the collapse of baryons at small scale, leading to more concentrated halos.

In this paper, we would like to extend the previous models for cosmic statistics to smaller scales, where baryons are likely to dominate the total mass distribution.
For that purpose, we use the results of a recent, high-resolution, cosmological simulation, featuring state-of-the-art galaxy formation physics, thanks to  the  Horizon collaboration\footnotemark[1].
The simulation was performed on the MareNostrum supercomputer at the Barcelona Supercomputer Centre using the \RAMSES \ code \citep{Teyssier2002}, including a detailed treatment of metal--dependent gas cooling, UV heating, star formation, supernovae feedback and metal enrichment. 
\footnotetext[1]{\url{http://www.projet-horizon.fr}}

The simulation data are compared to the analytical prediction of a \emph{modified halo model}, taking into account small scale baryons physics in an ad-hoc way by adding to the halo mass profile a small baryonic component, modelled as an exponential disk with mass fraction and scale length as the only 2 additional free parameters.
This approach, which modifies the shape of the halo profile, is in essence similar to the one of \citet{White2004}, which we use as a starting point for our theoretical model to compare against our numerical simulation.
%This model is based on the standard disk formation scenario of \citet{Mo1998} in a $\Lambda$CDM universe \citep[see also][]{Somerville2008}.
In contrast to previous studies, we also compute the effect of baryons on the \emph{skewness} of the mass distribution.
It has been shown that measuring the third moment of the cosmic shear is of paramount importance, since it can break the degeneracy in the cosmological parameters estimation based on the power spectrum alone, and reduce the corresponding error bars by a factor of 2 \citep{Bernardeau1997, Jain2000, Takada2000}.
Using only the two additional parameters of the model, we can fit the simulation data with great accuracy, for both the power spectrum and the skewness.
This has important consequences for future weak lensing surveys, since the disk parameters of the model could be fitted together with the cosmological parameters, promoting baryons physics from a mere systematic effect to an additional probe of the underlying cosmological model. Within the modified concentration model of \citet{Rudd2008}, statistical bias effects have been studied by \citet{Zentner2008}, and further by \citet{Hearin2009} in the context of the test of general relativity by weak lensing surveys.

\section{Statistical analysis of the MareNostrum simulation}

\subsection{The MareNostrum simulation}

We have performed a cosmological simulation of unprecedented scale, using 2048 processors of the MareNostrum computer installed at the Barcelona Supercomputing Centre in Spain.
We have used intensively the AMR code \RAMSES \ \citep{Teyssier2002} for 4 weeks dispatched over one full year.
This effort is part of a consortium between the Horizon project\footnotemark[1] in France and the MareNostrum galaxy formation project in Spain\footnotemark[2].
The originality of this project relies on using a lot of (if not all) physical ingredients that are part of the current theory of galaxy formation, and at the same time cover a large enough volume to provide a fair sample of the universe, especially at redshifts above 1.
\footnotetext[2]{\url{http://astro.ft.uam.es/\~marenostrum}}

The physical processes we have included in our simulation are described now in more detail.
We have considered metal-dependent cooling and UV heating using the Hardt and Madau background model \citep[see][]{Ocvirk2008}.
We have incorporated a simple model of supernova feedback and metal enrichment using the implementation described in \citet{Dubois2008}.
For high-density regions, we have considered a polytropic equation of state to model the complex, multi-phase and turbulent structure of the ISM \citep{Yepes1997,Springel2003} in a simplified form \citep[see][]{Schaye2008, Dubois2008}: the ISM is defined as a gas with a density above $n_0 \approx 0.1~\mathrm{H}/\mathrm{cc}$.
Star formation has also been included, for ISM gas only ($n_H > n_0$), by spawning star particles at a rate consistent with the Kennicutt law derived from local observations of star-forming galaxies.
In more mathematical terms, we have $\dot{\rho_\star} = \rho_{\textrm{gas}}/t_\star$ where $t_\star = (n_H/n_0)^{-1/2} t_0$ and $t_0 = 8~\textrm{Gyr}$.
Recast in units of the local free-fall time, this corresponds to a star formation efficiency of 5\%.
The simulation was started with a base grid of $1024^3$ cells and the same number of dark matter particles, and the grid was progressively refined, on a cell-by-cell basis, when the local number of particles exceeded 10.
A similar criterion was used for the gas, implementing what is called a quasi-Lagrangian refinement strategy.
Five additional levels of refinement were considered, providing a resolution between $1$ and $2 \kpc$ \emph{physical} at all times.
In this way, our spatial resolution is consistent with the angular resolution used to derive the Kennicutt law from observations.
On the other hand, we are not in a position to resolve the scale height of thin cold disks so that the detailed galactic dynamics might be affected by resolution effects.

The simulation was run for a $\Lambda$CDM universe with $\Omega_m = 0.3$, $\Omega_\Lambda = 0.7$, $\Omega_b = 0.045$, $H_0 = 70$ km/s/Mpc, $\sigma_8 = 0.9$ in a periodic box of $50 \Mpcoverh$.
Our dark matter particle mass ($m_p \approx 8 \times 10^6 \Msunoverh$), our spatial resolution (1 kpc \emph{physical}) and our box size make this simulation ideally suited to study the formation of galaxies within dark matter halos, from dwarf-- to Milky-Way--sized objects at high redshift.
For large galaxies, we can nicely resolve the radial extent of the disk (not its vertical extent), while for small galaxies, we can resolve the gravitational contraction of the cooling gas, but barely the final disk.
The simulation was stopped at redshift $z \approx 1.5$ because we ran out of allocated time.
The total number of star particles at the end of the simulation was above $2 \times 10^8$, and the total number of AMR cells was above $5 \times 10^9$.

To quantify the effects of baryons on statistical properties of the mass distribution, the MareNostrum run, which includes dissipative physics, was re-run from the same initial conditions with baryon mass converted to dark matter.
This latter dark matter only (henceforth DMO) simulation serves as a reference for statistical quantities without the presence of dissipative physics.
Both runs were carried out up to redshift $z=2$, which we will consider in the rest of this paper.
It is already late enough to witness interesting structures such as well-formed galaxy
disks.

\subsection{One-point statistics}

Meaningful statistics of the density field can be extracted from different statistical quantities, such as the $n$-point correlation functions, the density PDF, or the one-point cumulants.
By far, the easiest quantities to measure are the one-point moments $S_p(R)$, i.e. the $p$-th order cumulant of the smoothed density field as a function of the smoothing scale $R$.
The $S_p$ parameters have also been studied extensively, whether from a theoretical standpoint \citep{Balian1989,Szapudi1993}, in the perturbation theory framework \citep{Bernardeau1994}, or in simulations and observations \citep[see, e.g.,][]{Colombi2000, Marinoni2008}.
For the $50 \Mpcoverh$ box of MareNostrum, we have computed the statistics for scales ranging from $15 \kpcoverh$ to $2 \Mpcoverh$.

With weak lensing applications in mind, we are primarily interested in the total mass statistics.
In the case of the dissipative simulation, this requires a consistent treatment of both dark matter particles and gas cells.

The total local density in the dissipative simulation can be written
\begin{equation}
  \label{eq:rho_tot}
  \rho_{\textrm{tot}} = \rho_g + \rho_{\textrm{DM}} + \rho_s,
\end{equation}
where $\rho_g$, $\rho_{\textrm{DM}}$ and $\rho_s$ are the local gas, dark matter and star densities respectively.
However, because of the different nature of the gas (which is a continuous cell-based quantity), and stars and CDM (which are modelled as collisionless particles), the three matter components should be treated separately.
One could simply evaluate $\rho_{\textrm{DM}}$ and $\rho_s$ by binning the particles into cells to obtain a local estimate of the densities, and then simply calculate $\rho_{\textrm{tot}}$ as in Eq. \ref{eq:rho_tot} and computing its moments.
However, as we discuss below, the discrete nature of particles mandates a special treatment, and we chose instead to compute the moments and cross-correlations of the different species separately, and then reconstruct the moments of the total density field as we now describe.

Obtaining the moments of the gas distribution involves no theoretical difficulty.
The gas density of the whole simulation box is mapped onto a $2048^3$ grid from
the AMR cells using a donnercell scheme, where the resulting value in each cartesian cell is directly copied from the finest AMR cell covering it.
To determine the moments of the smoothed gas density field for a given comoving smoothing radius $R$, we compute the average of the density over cubic regions of volume $\frac{4}{3} \pi R^3$.
We restrict ourselves to values of $R$ corresponding to smoothing boxes which span an integer number of base grid cells.
The average densities in such cubic regions are computed using a fast convolution algorithm \citep[see e.g.][]{Blaizot2006}, and the moments over all such regions are then evaluated.
Since the simulation directly provides the continuous gas density $\rho_g$, this prescription yields unbiased estimates of the moments of the gas distribution.

\newcommand{\factmom}[2]{\ensuremath{\left({#1}\right)_{#2}}}
\newcommand{\avg}[1]{\ensuremath{\left<{#1}\right>}}
\newcommand{\pavg}[1]{\ensuremath{{\avg{#1}}_P}}

Particles require a somewhat more careful treatment.
The statistics of particle distributions are readily studied using a counts-in-cells analysis \citep[see for example][]{Balian1989,Bouchet1992,Sheth1996}.
The idea is to count particles within the same cubic cells of scale $R$ used for the smoothing of the gas density.
The particle counting is implemented by first binning the particles into the base grid using a nearest grid point (NGP) scheme \citep{Hockney1981}, and then counting particles in the cubic regions, again by using fast convolution.
This is indeed equivalent to computing a local particle density by NGP, and then performing the $R$-scale smoothing.
In this case, however, simply computing the moments of the resulting data will introduce shot noise effects \citep{Szapudi1993,Bernardeau2002}.

% Local Poisson sampling hypothesis for a continuous field
\newcommand{\smoothed}[1]{\tilde{#1}}
It is possible to correct for these effects using \emph{factorial moments}.
Let us consider a continuous field $\rho$ sampled by a finite collection of particles.
Given a cell of size $R$ and volume $v=R^3$, we call $\smoothed{\rho} = \frac{1}{v} \int_v \rho(\vect{x}) \mathrm{d}^3\vect{x}$ the average value of $\rho$ over the cell.
Equivalently, $\smoothed{\rho}$ is the value of $\rho$ smoothed at scale $R$ at the centre of the cell.
The local Poisson sampling hypothesis \citep[see e.g.][]{Bernardeau2002} states that the distribution of the number $N$ of particles in the cell follows a Poisson probability mass function of mean ${\tilde{\rho}v}/{m_p}$, where $m_p$ is the mass of a single particle.
%
% Factorial moments estimate the moments of the smoothed field
Letting
\begin{equation}
  \factmom{N}{k} \equiv N(N-1)\cdots(N-k+1),
  \label{eq:factdef}
\end{equation}
the factorial moments are defined as
\begin{equation}
  F_k \equiv \pavg{\factmom{N}{k}} = \pavg{N(N-1)\cdots(N-k+1)},
  \label{eq:factmoments}
\end{equation}
where $N$ is the cell particle count, and $\pavg{\dots}$ denotes the average over the Poisson distribution of the particle sampling.
It has been shown \citep{Szapudi1993} that the $F_k$ yield unbiased estimators of the moments of the underlying smoothed density field $\smoothed{\rho}$ at the scale of the cell size, in the sense that
\begin{equation}
  \smoothed{\rho}^k
  = \left(\frac{m_p}{v}\right)^k F_k
  = \left(\frac{m_p}{v}\right)^k \pavg{\factmom{N}{k}}.
  \label{eq:partmoments}
\end{equation}

% Computing the correlations of the smoothed fields
Let us now consider the density fields smoothed at scale $R$ for the gas, dark matter and stars, $\rho_g$, $\rho_{\mathrm{DM}}$ and $\rho_s$ respectively.
For the sake of readability, we shall drop the tilde notation, and the density fields are to be understood as smoothed at the scale $R$ in the rest of this section.
Using Eq. \ref{eq:rho_tot}, we can express the moments of the smoothed total density field $\rho_{\textrm{tot}}$ as a function of the moments and correlations of the individual species using the multinomial theorem:
\begin{equation}
  \label{eq:multinomial}
  \avg{\rho_{\textrm{tot}}^k} = \sum_{k_1+k_2+k_3 \leq k} {k \choose k_1, k_2, k_3}
  \avg{ \rho_g^{k_1} \rho_{\textrm{DM}}^{k_2} \rho_s^{k_3} }.
\end{equation}
In this equation, \avg{\cdots} denotes the ensemble average over all realisations of the underlying density fields, not to be confused with the average \pavg{\cdots} over particle samplings for a fixed realisation of $\rho$ introduced in equation \ref{eq:factmoments}.

Provided we can compute all correlations of the form $\avg{ \rho_g^{k_1} \rho_{\textrm{DM}}^{k_2} \rho_s^{k_3} }$, we are now in position to reconstruct the moments of the smoothed total density field.
Under the local Poisson sampling hypothesis, one can deduce from Eq. \ref{eq:factmoments} the identity:
\begin{equation}
  \avg{\rho_g^{k_1} \rho_{\textrm{DM}}^{k_2} \rho_s^{k_3}} = \left(\frac{m_{\textrm{DM}}}{v}\right)^{k_2} \left(\frac{m_s}{v}\right)^{k_3}
  \avg{\rho_g^{k_1} \left(N_{\textrm{DM}}\right)_{k_2} \left(N_s\right)_{k_3}},
  \label{eq:prescription}
\end{equation}
which involves the definition \eqref{eq:factdef}, and where $m_{\textrm{DM}}$ and $m_s$ are the dark matter and star particle masses.
Since the Poisson processes of the counts-in-cells for the different particle species are independent of each other, Eq. \ref{eq:prescription} involves no approximation, even though the underlying fields $\rho_{\mathrm{DM}}$ and $\rho_s$ are correlated.

From the moments \eqref{eq:multinomial}, we can straightforwardly compute the moments of the total matter overdensity $\avg{\delta_{\textrm{tot}}^k} = \avg{\left( \rho_{\mathrm{tot}}/\bar{\rho}_{\mathrm{tot}} - 1 \right)^k}$ from the binomial theorem.

We can finally compute the $S_k$ parameters, which are defined as
\begin{equation}
  S_k(R) \equiv \frac{\avg{\delta^k(R)}_c}{\avg{\delta^2(R)}_c^{k-1}},
\end{equation}
where the $c$ subscripts denote the connected moments of the smoothed density field, whose generating function is the logarithm of the generating function of the $\avg{\delta^k}$.

\subsection{Power spectrum}

Because of the particular significance of the 3D total matter power spectrum $P(k)$ in the convergence power spectrum, we have also measured $P(k)$ in the dissipative and DMO simulations, in addition to the one-point statistics.
The variance of the total matter density field smoothed at scale $R$ can be expressed as:
\begin{equation}
  \sigma^2(R) = \frac{1}{2\pi^2}
	\int \frac{\dd k}{k} k^3 P(k) \left| W(k R) \right|^2,
  \label{eq:sigma2}
\end{equation}
where $W$ is the Fourier transform of a spherical top-hat window function with volume unity:
\begin{equation}
  W(x) = \frac{3}{x^3} \left( \sin x - x \cos x \right).
\end{equation}
Various sophisticated techniques for estimating the power spectrum have been proposed, especially for correcting mass assignment and sampling effects \citep{Jing2005,Cui2008,Colombi2009}.
Since the 2-point correlation function (or, equivalently, the power spectrum) is not our primary interest in this paper, we have settled for a simple method which we expect to yield reasonable results, even if not as accurate as our one-point moments measurements.

The gas and particles densities were mapped onto a $2048^3$ base grid and added up, using donnercell for the gas and NGP binning for the dark matter particles.
The spectrum is computed using FFT folding \citep[see][]{Jenkins1998,Colombi2009} and corrected for the NGP convolution and shot noise bias effects \citep{Hockney1981}.

\subsection{Results}

\begin{figure}
\includegraphics[width=\columnwidth]{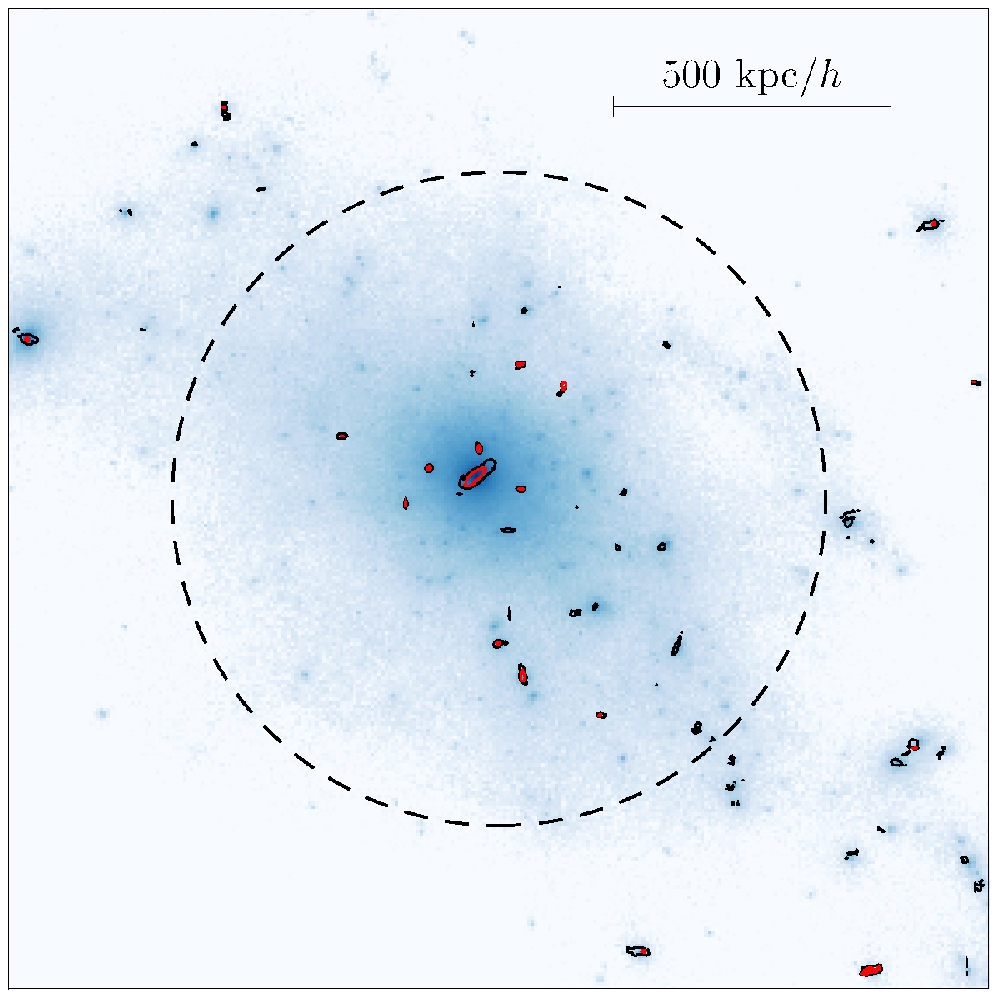}
\caption{Map of the projected dark matter density of one of the largest halos in the MareNostrum simulation ($\Rvir = 0.59 \Mpcoverh$ comoving shown as the dashed circle) at $z=2$.
The contours represent isovalues of the baryon to dark matter density ratio $\rho_\textrm{bar} / \rho_\textrm{CDM}$.
The outer black contours correspond to $\rho_{\textrm{bar}} = 0.1 \rho_{\textrm{CDM}}$, while the inner red contours delimit equal densities regions.
The total matter density is baryon-dominated at small scales well within the halo core.
The bright central galaxy clearly stands out of the halo substructures, whose distribution within the halo remains mainly unaffected by the presence of baryons \citep[see][]{Weinberg2008}. }
\label{fig:halo}
\end{figure}

Because of cooling, the baryons will condense to form dense structures such as disks at the centre of dark matter halos.
Figure \ref{fig:halo} shows a density map of one of the biggest MareNostrum halos, together with contours of the density ratio $\rho_\textrm{bar} / \rho_\textrm{CDM}$.
The effect of cooling can be seen as dense baryon-dominated regions at the cores of the halos and halo substructures.

The small-scale baryonic features directly impact the density statistics at small scales: as the smoothing scale decreases, the disks become more and more apparent in the density PDF as peaks in the high-density regions.
We can expect this process to broaden the distribution, thereby increasing the variance, and as only the higher-density regions are affected, the skewness should also increase.

\begin{figure*}
\includegraphics[width=0.49\textwidth]{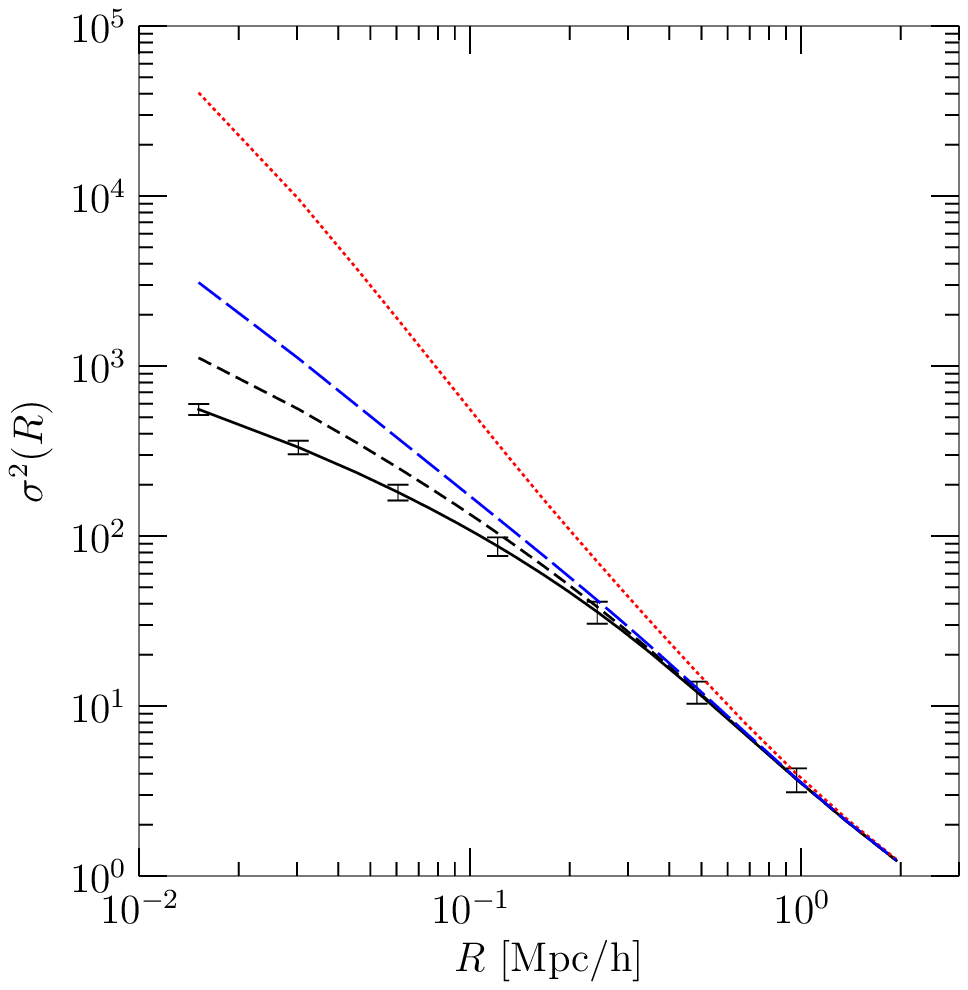}
\includegraphics[width=0.49\textwidth]{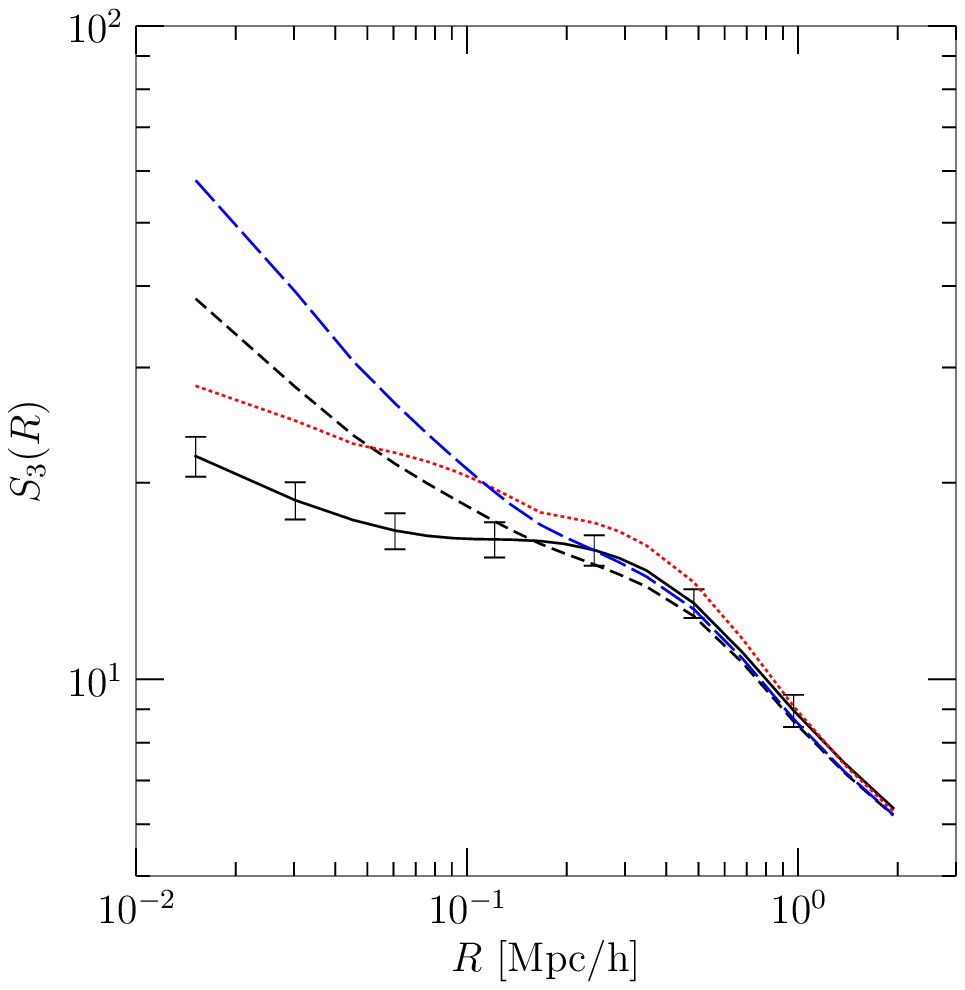}
\caption{Variance (left) and skewness (right) of the smoothed density field of different species at $z=2$, as a function of the smoothing scale in the MareNostrum dissipative and DMO simulations.
The solid line shows $\sigma^2$ and $S_3$ for dark matter in the DMO simulation, while the dashed and dotted lines correspond to the dissipative simulation: short dashes for dark matter, long dashes for total matter, and dots for baryons.
The error bars on the DMO data are one-sigma wide and determined by the subvolumes method as described in the text.}
\label{fig:mn_data}
\end{figure*}

The computed variance $\sigma^2$ and skewness $S_3$ from the MareNostrum dissipative and DMO simulations is presented on Fig. \ref{fig:mn_data}.
Comparing the DMO simulation (solid black) with the total matter in the dissipative run (blue dashes), we indeed note that the presence of baryonic physics dramatically amplifies both $\sigma^2$ and $S_3$ at small $R$.
At $k \sim 10 \hoverMpc$, the power spectrum boost reaches about 35\% (see Fig. \ref{fig:pk-boost}), most of which is caused by cold baryons (stars).
Because our study is carried out at $z=2$, precise comparisons with previous results of \citet{Jing2006,Rudd2008} are difficult.
Note however that we observe the same qualitative effects.
The variance plot on Fig. \ref{fig:mn_data} also demonstrates the presence of CDM adiabatic contraction \citep{Gnedin2004}.
As the gas cools down, its contraction drags the dark matter into local potential wells created by dense baryon clumps.
This effect results in a net condensation of the dark matter, whose effect on variance can be seen by comparing the DMO run (solid black curve) with CDM of the dissipative run (short-dashed curve).
Both observed boosts and dark matter contraction effects are well in accordance with the results presented in \citet{Weinberg2008}.

Because of the relatively small size of the MareNostrum simulation box, the results presented on Fig. \ref{fig:mn_data} are contaminated to some degree by cosmic variance and finite volume effects.
We have estimated those effects in the MareNostrum DMO simulation.
Note that the rigorous determination of error bars is beyond the scope of this article, and we do not expect baryons to modify those uncertainties significantly.

\begin{figure*}
\includegraphics[width=0.49\textwidth]{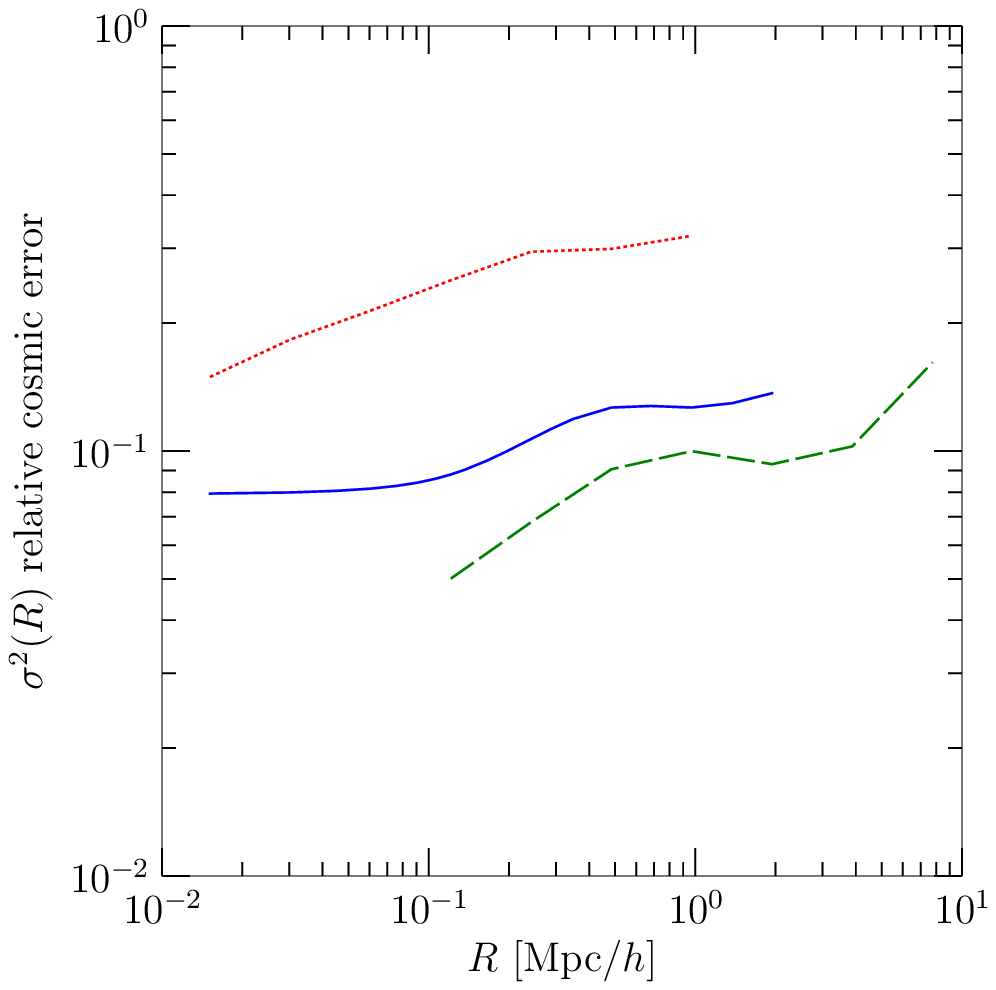}
\includegraphics[width=0.49\textwidth]{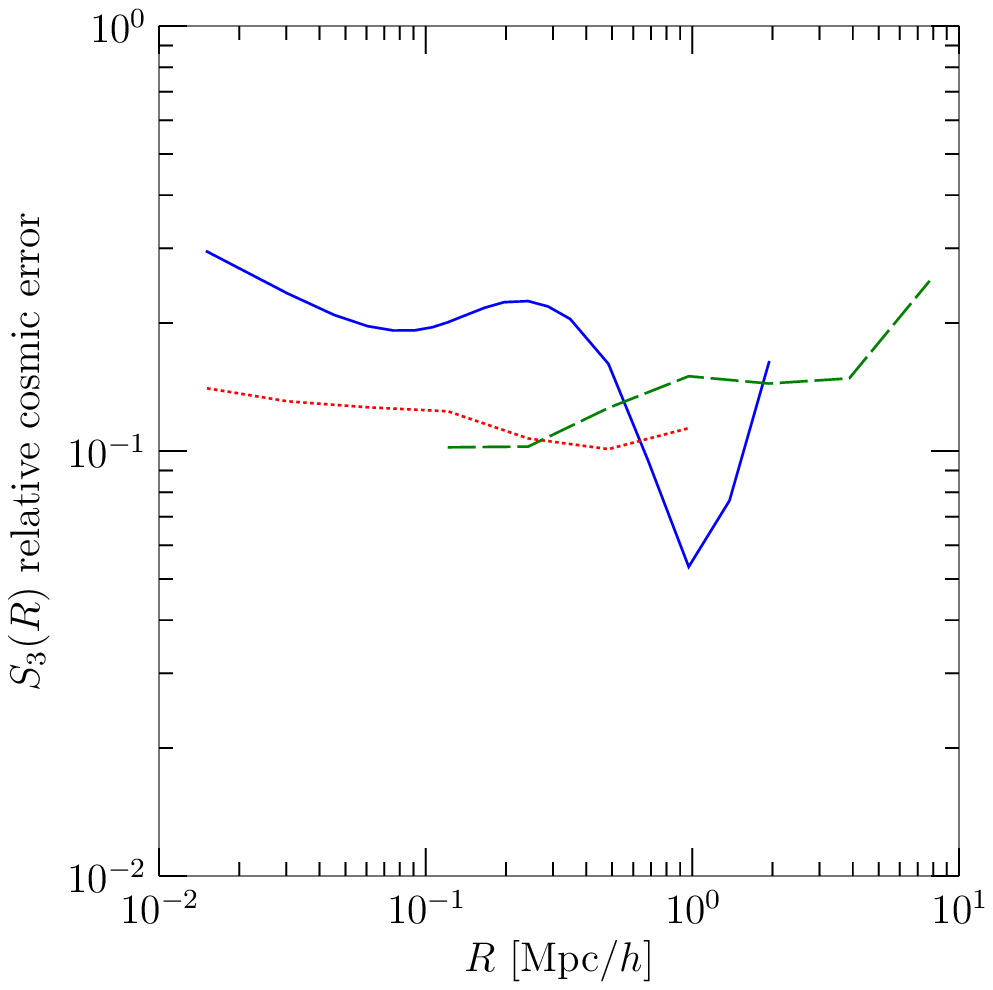}
\caption{Estimates for the relative cosmic errors $\Delta \sigma^2 / \sigma^2$ and $\Delta S_3 / S_3$ for each method described in the text.
The dashed curves correspond to the 10 ensemble simulations, the solid curves to results of the FORCE code, and the dotted curves to the subvolumes estimation.}
\label{fig:errs}
\end{figure*}

The cosmic variance and finite volume effects on the statistical quantities were sampled by three different independent methods.
% Sims
We have run a set of 10 smaller $256^3$ cosmological simulations up to $z=2$ with the same box size and power spectrum as the MareNostrum box, only with differing random phases.
The statistical quantities were then computed on each box, and the variance of those quantities over the 10 boxes were used as a first estimate of the MareNostrum cosmic variance effects.
While such ensemble simulations are easy to carry out, this method is known to underestimate the actual cosmic variance, as all the realisations of the initial density field are constrained: the total box matter density is fixed to the background matter density of the universe.
In addition, this method cannot be used to determine the variance at small scales because of the low resolution of the ensemble simulations.
Relative cosmic error derived from this set of simulations is presented on Fig. \ref{fig:errs} (dashed curve).
% FORCE
The FORCE code \citep[FORtran for Cosmic Errors][]{Colombi2001}, implements the results of \citet{Szapudi1999} and provides cosmic variance estimations given the values of the density cumulants.
The corresponding cosmic error, based on the MareNostrum DMO cumulants, is shown as the solid curves on Fig. \ref{fig:errs}.
This estimation relies on a perturbative expansion which breaks down when relative errors approach unity.
As the MareNostrum errors range from about 5\% to 30\%, the FORCE computation still holds, but the quality of the estimation is impacted, especially at small scales where the errors on high-order cumulants increase.
% Subvolumes
To confirm the FORCE results at small scales, we have studied the variance of the statistical quantities over a random sample of cubic subvolumes of size $\ell$.
Let $\epsilon_X(\ell, R) = \sqrt{\mathrm{var(X(R))}}/\bar{X}(R)$ be the relative cosmic error of a statistical quantity $X$ at scale $R$ defined on subvolumes of size $\ell$.
To obtain the cosmic variance of the whole simulation box (i.e., $\epsilon_X(L,R)$ for all $R$), we computed $\epsilon_X(\ell, R)$ for $\ell = L/8, L/16, L/32$ and extrapolated in $\ell$ to $\ell=L$ assuming the power-law form $\epsilon_X(\ell, R) = \epsilon_X(L,R) (\ell/L)^\eta$.
This last estimation of the error is represented on Figure \ref{fig:errs} in dotted lines.
None of these methods ensures accurate determination of the errors over the whole range of scales, however, they paint a clear picture of cosmic variance in the MareNostrum simulations.
As can be seen on Fig. \ref{fig:mn_data}, the observed boosts in $\sigma^2$ and $S_3$ are well above cosmic variance effects.
Note that scales comparable to the MareNostrum box size correspond to a patch of $z=0.5$ sky extending over about $4$ squared degrees.

For our present study, it is important to notice, however, that since both the DMO and dissipative runs have been performed using the same set of random phases for the initial conditions, they suffer from the same such effects.
As a consequence, the corresponding errors in the two runs are strongly correlated, and ratios of statistical quantities such as $\sigma^2_{\textrm{tot}} / \sigma^2_{\textrm{DMO}}$ are mostly devoid of finite volume contamination. 
For the rest of this paper, we will therefore only consider such amplification ratios (or ``boosts'') for the variance and skewness of the total matter in the dissipative run with respect to the dark matter of the DMO run.

\section{A modified halo model for baryons}

\subsection{The halo model}

The halo model provides a well-tested and flexible framework for the study of the properties of matter distribution in non-linear stages of gravitational collapse.
While first studied in the context of the galaxy distribution \citep{Neyman1952}, it has become a full-fledged and now mature tool for cosmological statistics through significant contributions and improvements to its various ingredients.

Attempting to reproduce the effect of baryons on the boost factors of variance and skewness requires us to model both the DMO and dissipative matter distributions.
\citet{Rudd2008} have shown that modifying the halo concentration relation can account for the variance amplification at scales $k \lesssim 20 \hoverMpc$.
In this paper, we will use a standard halo model to describe the dark matter of the DMO run.
We base our halo profile for the total mass on the DMO halo model, but instead of modifying only $c(M)$, we propose to also modify the halo profile itself.
As discussed previously, the quantity of interest is the boost of the statistics (i.e. the amplification of the variance and skewness witnessed on the total matter halo model with respect to the reference halo model).
We now briefly describe the different key ingredients which take part in the computation of $\sigma^2(R)$ and $S_3(R)$ in the halo model.

Statistical description of the density field as a set of virialized halos requires the specification of the mass distribution of the halos (the mass function), their density profiles and associated mass parametrisation, and a model for halo-halo correlations.

A simple model for the halo mass function was given by \citet{Press1974} based on the spherical collapse model.
Since then, there has been more convincing derivations of the Press-Schechter result, as well as  attempts to take into account non-symmetric collapses and tidal effects \citep{Bond1991,Audit1997,Sheth2001,Sheth2002}.
These studies resulted in other parametrizations, such as the Sheth-Tormen mass function \citep{Sheth1999}.

In this study, we use the Sheth-Tormen prescription for the halo mass function, as it turns out to be a better fit to our simulations than the Press-Schechter form.
In normalized units, the Sheth-Tormen mass function reads:
\begin{align}
  \begin{split}
	\label{eq:ST}
	\frac{m}{\bar{\rho}} n(m) \dd m = {} & A(p) \sqrt{\frac{2q}{\pi}}
	  \left(1 + {\left(q\nu^2\right)}^{-p}\right) \\
	  & \times \nu \exp \left( -\frac{q\nu^2}{2} \right)
	  \frac{\dd \nu}{\nu},
  \end{split}
\end{align}
where $\nu \equiv \delta_{c}/\sigma$. $\delta_{c} \approx 1.68$ is the collapse density threshold in the spherical collapse model, and $p \approx 0.3, A(p) \approx 0.322, q \approx 0.75$.

We have also introduced a mass cutoff in the halo mass function to account for the small box size of the MareNostrum simulation.
The value of the cutoff is chosen slightly above the mass of the biggest halo found in the simulation, which is around $5. 10^{13} \Msun$.
While the cutoff has little effect on the variance as computed by the halo model, the skewness drops significantly at large scales, resulting in a better fit against the measured $S_3$.
This is not surprising since high-order moments at large scale are sensitive to rare events (such as massive halos, e.g. \citealt{Colombi1994}), which are not well sampled by the MareNostrum box.

For the DMO model halo profile, we use the standard NFW form \citep{Navarro1997}:
\begin{equation}
  u_{\textrm{NFW}}(r|M) \propto x^{-1} \left( 1 +  x \right)^{-2},
	\quad x \equiv \frac{c(M)r}{\Rvir}
  \label{eq:nfw}
\end{equation}
and $u(r|M)$ is normalized such that $\int u(r|M)\dd^3\vect{r}=1$.
Our halo virial radius $\Rvir$ is defined such that, for a halo of mass $M$, we have $M = \frac{4}{3}\pi \bar{\rho} \Rvir^3 \Delta$, with $\Delta=200$.
Note that the mass $M$ of a halo is related to its comoving Lagrangian size $R$ by $M =  \frac{4}{3}\pi \bar{\rho} R^3$.
The NFW model has proved to fit numerical dark matter profiles over a large range of masses and radii with some dispersion in the concentration parameter $c(M)$ \citep{Kravtsov1998,Jing2000}.
The central logarithmic slope of dark matter profiles, which is $-1$ in the case of NFW, is currently debated (see \citealt{Fukushige1997,Moore1998}, and more recently \citealt{Springel2008,Stadel2008}).
Note however that in the presence of dissipative physics, baryons are likely to affect the inner slope.

The concentration parameter $c(M)$ is parametrized in our model according to the result of \citet{Bullock1999}:
\begin{equation}
  c(M,z) = \frac{c_0}{1+z} \left( \frac{M}{M_*} \right)^b,
  \quad c_0 \approx 9, \quad b \approx -0.13.
  \label{eq:c}
\end{equation}
This power-law model has been found to be a good fit to numerical simulations also for dark energy cosmologies \citep{Dolag2004}.

Following \citet{Scherrer1991}, we can express the density 2-point correlation function $\xi(r)$ as:
\begin{equation}
  \xi(r) = \xi_{1h}(r) + \xi_{hh}(r),
  \label{eq:halo_xi}
\end{equation}
where $\xi_{1h}$ represents the contribution to the correlation function from mass within the same halo, and $\xi_{hh}$ contains the contribution from mass located in different halos.

$\xi_{1h}$ is essentially the autocorrelation of the halo profile, and its contribution to the power spectrum is:
\begin{equation}
  P_{1h}(k) = \int n(m) \left(\frac{m}{\bar{\rho}} \right)^2 \left| u\left(k|m\right) \right|^2 \mathrm{d} m,
\end{equation}
where $u(k|m)$ is the Fourier transform of the halo profile for a halo of mass $m$.
We compute the halo-halo contribution following \citet{Mo1996} and its subsequent extensions \citep{Mo1997,Sheth1999a,Sheth1999}.
Assuming deterministic biasing on large scales, we can write the $\xi_{hh}$ contribution from two halos of masses $M_1$ and $M_2$ as:
\begin{align}
  \begin{split}
	\xi_{hh}(r|M_1,M_2) &= b(M_1) b(M_2) \xi(r) \\
	& \approx  b(M_1)b(M_2)\xi_{\textrm{lin}}(r),
  \end{split}
\end{align}
where $\xi_{\textrm{lin}}$ is the matter correlation function from linear theory.
This prescription is accurate at large scales, and consistent with the choice of mass function provided the bias $b(M)$ is computed from $f(\nu)$ as prescribed in \citet{Mo1997}.

Now in possession of a halo model for $\xi(r)$ (and therefore its Fourier transform $P(k)$), we can evaluate $\sigma^2(R)$ using Eq. \ref{eq:sigma2}.

\subsection{Skewness in the halo model}

While in principle the halo model ingredients presented so far fully determine the statistics of the density field, additional work is needed to extract $S_3(R)$.

At large enough scales, the one-point statistics $S_k$ may be computed using perturbation theory \citep{Fry1984a,Juszkiewicz1993,Bernardeau1994,Bernardeau2002}, which yields
\begin{equation}
  S_3^{\mathrm{PT}} = \frac{34}{7} + \gamma,
\end{equation}
where $\gamma = \mathrm{d} \ln \sigma^2(R) / \mathrm{d} \ln R$.
However, in the MareNostrum simulation at $z=2$, PT is only expected to be valid at scales greater than $\sim 1 \Mpcoverh$.
A first interesting refinement taking discrete halos into account is the Poisson cluster model, where halo-halo correlations are neglected and profiles are assumed to be point-like \citep{Sheth1996}.
Halo profiles, however, are responsible for most of the behaviour of small-scales statistics, and thus neither perturbation theory and the point-cluster model are appropriate for our study.

Fortunately, the full computation of the higher-order cumulants $S_k$ in the halo model was developed in \citet{Scoccimarro2001}.
Following the authors, we define:
\begin{align}
  \overline{u^m}(R,\nu) \equiv  &
	\int \frac{k^2 \mathrm{d}k}{2\pi^2} \left[ u(k|\nu) \right]^m \left| W(kR) \right|^2 \\
  A_{i,j}(R) \equiv & \int \mathrm{d} \nu f(\nu) b_i(\nu)
	\overline{u^2}(R,\nu) \left[ \overline{u}(R,\nu) \right]^j
	\left( \frac{M}{\bar{\rho}} \right)^{j+1},
\end{align}
where $R$ is such that $\delta_{c}/\sigma(R)=\nu$.
In these notations, the third cumulant of the density field in the halo model writes
\begin{equation}
  \left< \delta^3 \right>_c = S_3^{\mathrm{PT}} \sigma^4_{\mathrm{lin}}
	+ 3 \sigma^2_{\mathrm{lin}} A_{1,0} + A_{0,1}.
\end{equation}

\subsection{Halo model results}

We have tested some families of halo profiles to attempt to reproduce the observed effect of baryons on the statistics of the density field.
The reference halo model for the DMO simulation is based on a NFW profile with the commonly used $c(M)$ relationship of \citet{Bullock1999} as written in Eq. \ref{eq:c}.

\begin{figure}
\includegraphics[width=\columnwidth]{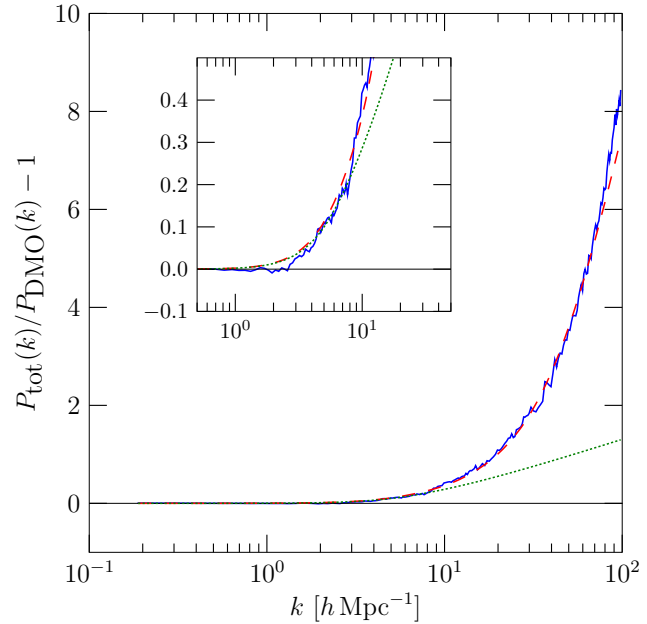}
\caption{Relative power spectrum amplification due to baryons at $z=2$.
The solid curve is the measured power spectrum, the dotted curve is a NFW profile with $c_0=20$, $b=-0.15$, and the dashed curve is the halo model with our composite halo profile.}
\label{fig:pk-boost}
\end{figure}

\begin{figure*}
\includegraphics[width=0.49\textwidth]{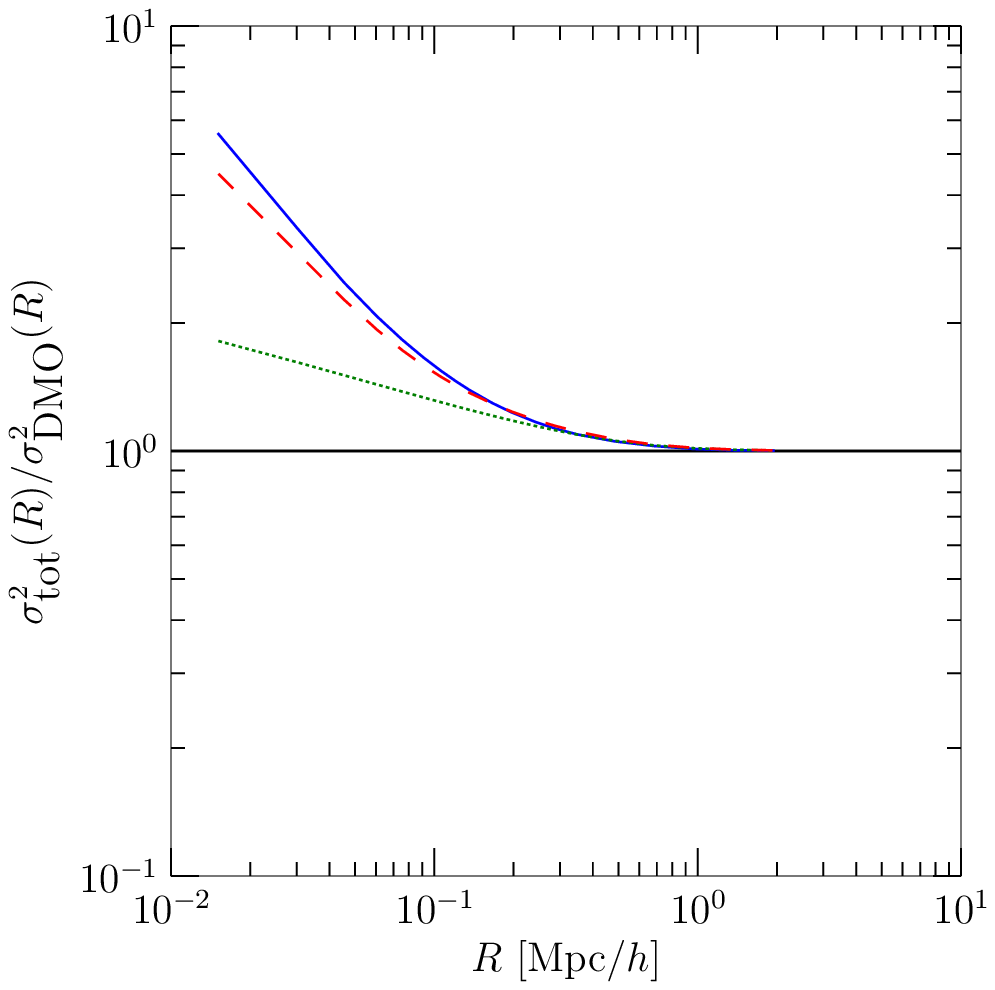}
\includegraphics[width=0.49\textwidth]{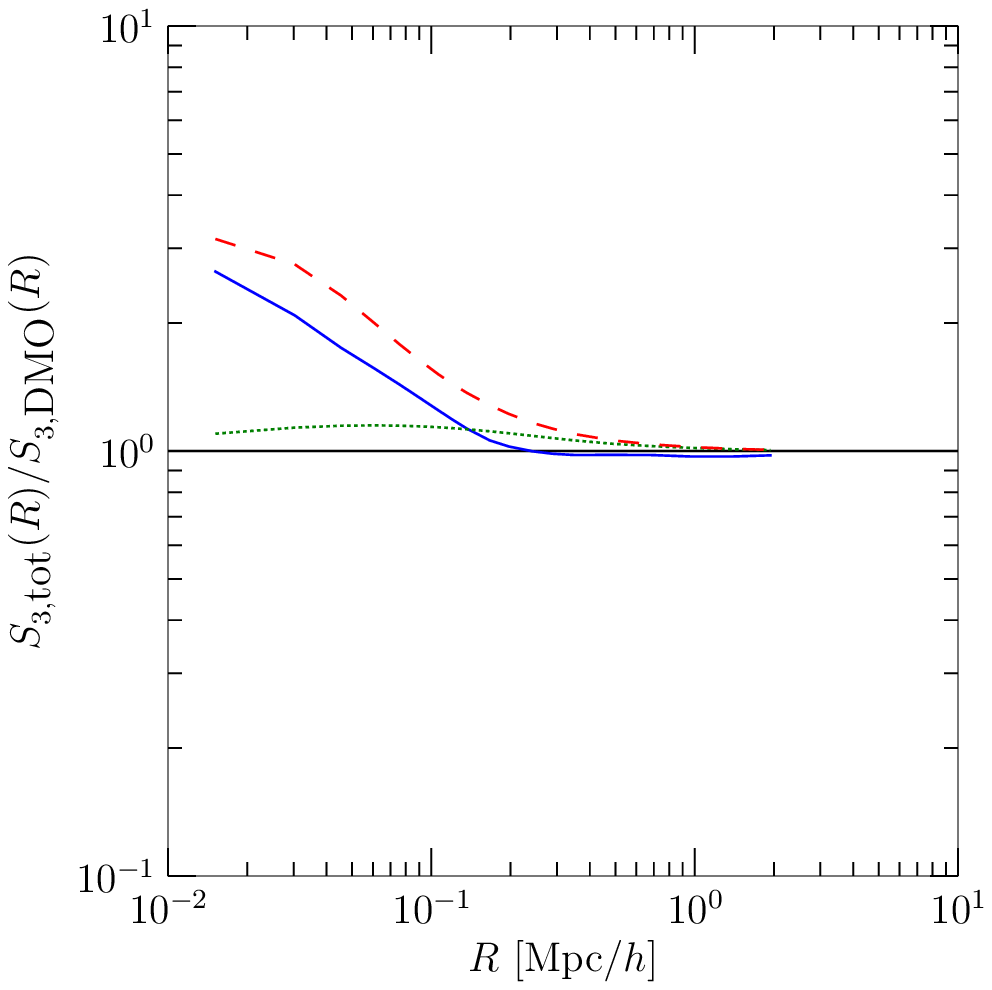}
\caption{Effect of baryons on the variance and the skewness $S_3$ boost factors, as measured on the MareNostrum simulation (solid curve) and modelled by a NFW profile with $c_0=20$, $b=-0.15$ (dotted curve), and the composite profile (dashed curve).}
\label{fig:stats-boost}
\end{figure*}

As suggested by previous numerical studies \citep{Rudd2008}, an increase in $c_0$ and a steeper concentration slope $b$ are expected to reproduce -- at least partially -- the increase in power at small scales due to baryonic physics and radiative processes in particular.
We have accordingly tried to adjust the concentration parameters with a NFW profile to obtain a good match for both the variance and skewness at small scales.
The power spectrum, variance and skewness boosts for a NFW-based model with parameters comparable to \citet{Rudd2008} ($c_0=20, b=-0.15$) are presented as the dotted curves on Fig. \ref{fig:pk-boost} and \ref{fig:stats-boost}.
This model reproduces the MareNostrum variance and power spectrum boosts down to a scale of about $0.5 \Mpcoverh$.
At smaller scales however, the halo model underestimates the variance amplification.
A large part of this discrepancy is likely due to the difference in the simulation codes and physical modelling between the two studies.
Note however that the skewness $S_3$ of this halo model lacks much of the measured small-scale amplification, as can be seen on Fig. \ref{fig:stats-boost}.
The distinctive bend is also not reproduced at all, which suggests the profile form distributes matter too evenly across scales.

\begin{figure}
\includegraphics[width=\columnwidth]{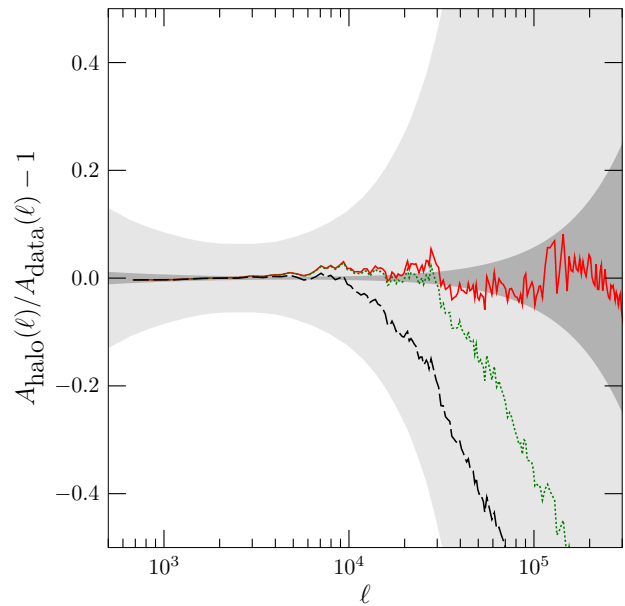}
\caption{Error on the amplification of the 3D total mass power spectrum at $z=2$ for the halo models represented as a function of the angular mode $\ell$ in the flat sky approximation.
The dashed curve is the reference DMO model (i.e. without any boost), the dotted curve is the pure NFW model with modified $c(M)$, and the solid curve is the composite halo model amplification.
The light and dark shaded areas are estimates of the expected experimental errors on $C_\ell$ for the CFHT Wide-field and LSST experiments, respectively.}

\label{fig:power-err}
\end{figure}

With the partial success of this profile, one might expect NFW profiles with higher concentrations to yield better fits.
It turns out however that reasonably fitting the variance boost at small scale requires very high values of $c_0$, exceeding $30$.
Such high values of the concentration parameter are too high to be accepted as physically meaningful.
Yet more importantly, while increasing $c_0$ will indeed boost the variance, it fails to reproduce at all the corresponding small-scale skewness amplification.
This can be seen on Fig. \ref{fig:stats-boost}, and the $S_3$ boost of a pure NFW halo model remains essentially flat for varying values of $c_0$, with a very weak dependence.

This leads us to believe that, while the NFW profile with adjusted concentration parameters has merits in modelling the variance amplification caused by dissipative physics, it can only paint a limited picture of the statistical properties of the density field in the presence of baryons.
As increasing $c_0$ essentially amounts to concentrating more matter within the central region of the halos, we naturally turn to other centrally-concentrated halo profiles.

One way to concentrate more matter within the centre region is by using families of profiles with steeper central cusps than NFW of the form:
\begin{equation}
  u_{\alpha}(x) = x^{-\alpha} (1+x)^{\alpha-3},
\end{equation}
where $\alpha=1$ yields an NFW profile.
We have tested this family of profiles on a wide range of values $1 \leq \alpha \leq 2.5$.
For each value of $\alpha$, we attempted to find an best-fitting value of $(c_0, b)$, again by exploring the parameter space.
It is interesting to note that high values of $\alpha$, in the range $[2.0, 2.15]$, produce to some extent both the $\sigma^2$ small-scale steepening and a strong $S_3$ amplification.
Isothermal ($\alpha=2$) profiles are known to be a good description of the total density in haloes hosting elliptical galaxies \citep[see e.g.][]{Gavazzi2007, VanDeVen2009}.
In the case of the MareNostrum simulation however, this property seems coincidental, as the simulated physics form no truly elliptical galaxy comparable to observations.
Moreover, the residuals of the best $\sigma^2$ and $S_3$ fits for such profiles cast doubt on the legitimacy of the analytical form $u_{\alpha}$ for the statistical analysis of the simulation.

\subsection{A modified halo profile}

A good candidate profile which is both centrally-concentrated and physically-motivated is a \emph{composite halo profile} \citep[see][]{White2004, Zhan2004}, parametrized by the dimensionless parameters $f_d$ and $\lambda_d$:
\begin{align}
  \begin{split}
	u_{f_d, \lambda_d} (r|M) = {} & \left(1-f_d\right) u_{\textrm{NFW}}(r|M) \\
	  & + f_d \, u_{\textrm{exp}, \lambda_d}\left(r | M\right),
	\label{eq:composite}
  \end{split}
\end{align}
where $u_{\textrm{exp},\lambda_d}$ is a spherically averaged exponential disk profile with length scale $r_d$ proportional to the halo's virial radius:
\begin{equation}
  u_{\textrm{exp},\lambda_d}(r|M) \propto \frac{\exp (-r/r_d)}{r/r_d},
  \quad r_d \equiv \lambda_d \Rvir.
\end{equation}
The dimensionless parameter $\lambda_d$ is essentially the spin parameter of the halo, and defines the disk scale $r_d$.
The profile $u_{f_d,\lambda_d}$ features a central $r^{-1}$ cusp and behaves like the NFW profile for radii bigger than the disk length scale $r_d$.
However, because of the profile normalization, it concentrates more mass within the central exponential than a pure NFW.
$u_{f_d,\lambda_d}$ can be seen as a halo profile concentrating a fraction $f_d$ of the mass within a central exponential disk profile, and the remaining $1-f_d$ in a standard NFW component.

This form of composite profile is physically motivated.
The total mass distribution in group-sized halos is known to be well described by a halo component and a concentrated component corresponding to the bright central galaxy \citep[see, e.g.,][]{Dubinski1998}.
The presence of baryons does not fundamentally change the diffuse halo component:
the distribution of satellite galaxies within halos is very similar to the halo occupation distribution of dark matter substructures in pure $N$-body simulations \citep[see][]{Weinberg2008}.
This suggests keeping a NFW profile to account for the dark matter, diffuse gas and halo substructures, while introducing a spiked central component mimicking the bright central galaxy's disk.
We may expect this NFW profile to be more concentrated than in the dark matter only case, because of the adiabatic contraction of the CDM due to the presence of baryons.
This will therefore lead to an increase of $c_0$ in the $c(M)$ relationship of equation \eqref{eq:c}.
For the composite profile, $f_d$ is to be understood as the fraction of the total halo mass which resides in the galactic disk in the form of baryons.
As most formed galaxies found at $z=2$ in MareNostrum simulation are spirals, we restrict ourselves in this paper to an exponential disk profile for the central component.
We believe this form captures the essential features of the dense central baryonic regions which are important for the halo model.
It also places interesting constraints on the profile parameters $f_d$ and $\lambda_d$, as mass fractions and angular momenta of disks are well-studied, both theoretically and observationally.
We further assume both $\lambda_d$ and $f_d$ to be independent of halo mass.
The assumption that the disk size is a fixed fraction of $\Rvir$ corresponds to the singular isothermal sphere model of disk formation \citep[see][]{Mo1998}.
We postpone refinements of this model to future work.

Here again, we explored the $(f_d, \lambda_d)$ parameter space to find a good fit to the MareNostrum data.
Our best model has parameters:
\begin{equation}
  \left\{
	\begin{array}{rcl}
	  c_0  &=& 13.5 \\
	  b    &=& -0.15 \\
	  f_d  &=& 0.09 \\
	  \lambda_d  &=& 0.025
	\end{array}
  \right.
  \label{eq:bestparams}
\end{equation}
The corresponding power spectrum, variance and skewness boosts are represented on Fig. \ref{fig:pk-boost} and \ref{fig:stats-boost} as dashed curves.
This halo profile reproduces accurately both the measured $\sigma^2$ and $S_3$ amplifications down to the smallest scales.

\section{Discussion and conclusion}

With a base grid resolution and particle count of $1024^3$ and a box size of $50 \Mpcoverh$, the MareNostrum simulation resolves the length and mass scales of galactic disks while also providing a volume large enough for cosmological studies.
This makes it particularly suitable for the study of the effect of baryonic physics on cosmic statistics.
Such an intermediate box size, however, will be affected at both small and large scales by resolution and finite volume effects.

At very small scales, counts-in-cells measurements are expected to suffer from shot noise, as the density field is sampled by a finite number of particles.
This translates into both statistical variance and bias at small scales, if using naive statistical estimators for the moments of the density field.
Assuming particles trace the density field as a local Poisson process, it can be shown, however, that factorial moments defined in Eq. \ref{eq:factmoments} are unbiased estimators \citep{Szapudi1993,Bernardeau2002}.
We thus do not expect our measurements to be affected by Poisson noise at small scales.

On large scales, the results will be contaminated by cosmic variance and finite volume effects.
In cosmological simulations, statistical quantities are usually computed by taking the spatial average -- instead of ensemble average -- of local quantities over the single simulated volume.
This prescription is only appropriate for scales corresponding to wavenumbers $k$ for which the simulation provides sufficient independent samples.
For a box of a given size $L$, the sampling of large scales $k$ with $\frac{2\pi}{k}$ approaching $L$ suffers from the decreasing number of independent modes.
The low number of modes at low wavenumbers introduces variance on large scale quantities, which is purely statistical in nature.
As presented in earlier, we have measured the cosmic variance of the whole MareNostrum box, and while conservative estimates for the errors range from 5\%--30\% depending on the statistic and estimation method, it is our understanding that cosmic variance does not fundamentally affect our result.
As previously mentioned, we have minimized the effect of cosmic errors on our conclusions by only considering ratios of statistical quantities from simulations run with the same set of random phases. 

We have shown that, although different halo profiles can describe variance amplification due to dissipative physics at small scale by merely modifying the
concentration parameter $c(M)$, the third moment $S_3$ introduces additional constraints on the inner profiles which cannot be reproduced by changing $c(M)$ alone.
The distinctive slope of $S_3(R)$ at small scales seems characteristic of a higher mass concentration towards the core than NFW.
Unsurprisingly, profiles with a core or relatively weak central density peaks do not describe well the effective total matter distribution in the presence of baryons.

Instead, we have found that using a superposition of a NFW profile and an exponential profile yields realistic variance amplification and $S_3$ gain for reasonable values of the concentration parameters $c_0$ and $b$, disk mass fraction $f_d$ and disk scale $\lambda_d$.
One should note that the values of the best-fitting $\lambda_d$ and $f_d$ parameters are in good agreement with the expected physical properties of the galaxies of the simulated MareNostrum universe.
The $f_d = 0.09$ value is quite compatible with observed and predicted baryon disk mass fractions \citep[see, e.g.,][]{Somerville2008}.

In this model, we chose not to introduce any mass or redshift dependence in $f_d$ and $\lambda_d$.
For the latter, this assumption is supported in part by the weak dependence on mass of $r_d/\Rvir$ \citep{Somerville2008}.
On the other hand, for $f_d$, a proper model should account for the variation of $M/L$ as a function of halo mass in real galaxies \citep[see, e.g.,][]{Yang2003}. We postpone this more elaborate approach to a future paper.

The modified $c_0$ and $b$ parameters of Eq. \ref{eq:bestparams} for the NFW profile correspond to a more concentrated CDM component than in the pure DMO model.
This is in accordance, both qualitatively and quantitatively, with the results of \citet{Rudd2008}.
It is interesting to note that the variance boost caused by the NFW component is of the same order of magnitude as the adiabatic contraction effect visible on Fig. \ref{fig:mn_data}, albeit slightly weaker.
This supports the idea that the composite halo profile concentrates a significant fraction of the halo's baryonic mass within the central disk, while the remaining halo gas essentially follows the NFW component which accounts for the cold dark matter.
This last CDM component ``feels'' the presence of the hot gas through the process of adiabatic contraction.

This suggests that both variance and skewness of the density field can be estimated at small scale within the framework of the halo model by using a composite halo profile.
While the halo model is a valuable tool for the study of theoretical power spectra, the accuracy requirements of precision cosmology are arguably too stringent to consider directly fitting cosmological parameters to observed cosmic statistics.
The halo model has merit however, as it allows one to study the dependence of cosmic shear with baryonic features such as galaxy disk masses and sizes.

\section*{Acknowledgements}
\AckText

\bibliography{cosmo}
\bsp
\label{lastpage}

\end{document}